\documentclass[%
 aip,
 amsmath,amssymb,
 reprint,%
]{revtex4-1}

\usepackage{graphicx}
\usepackage{dcolumn}
\usepackage{bm}

\usepackage[utf8]{inputenc}
\usepackage[T1]{fontenc}
\usepackage{mathptmx}
\usepackage{etoolbox}
\usepackage{multirow}

\makeatletter
\def\@email#1#2{%
 \endgroup
 \patchcmd{\titleblock@produce}
  {\frontmatter@RRAPformat}
  {\frontmatter@RRAPformat{\produce@RRAP{*#1\href{mailto:#2}{#2}}}\frontmatter@RRAPformat}
  {}{}
}%
\makeatother
\begin{document}

\preprint{AIP/123-QED}

\title[Predicting Non-Ideal Effects from the Diaphragm Opening Process in Shock Tubes]{Predicting Non-Ideal Effects from the Diaphragm Opening Process in Shock Tubes}

\author{Janardhanraj Subburaj}
  \email{janardhanraj.subburaj@kaust.edu.sa}
\author{Miguel Figueroa-Labastida}%
\author{Aamir Farooq}
\affiliation{Mechanical Engineering, Physical Science and Engineering Division, King Abdullah University of Science and Technology (KAUST), Thuwal 23955, Kingdom of Saudi Arabia.}

\date{\today}

\begin{abstract}
Shock tubes are instrumental in studying high-temperature kinetics and simulating high-speed flows. They swiftly elevate the thermodynamic conditions of test gases, making them ideal for examining rapid chemical reactions and generating high-enthalpy flows for aerodynamic research. However, non-ideal effects, stemming from factors like diaphragm opening processes and viscous effects, can significantly influence thermodynamic conditions behind the shock wave. This study investigates the impact of various diaphragm opening patterns on the shock parameters near the driven section end-wall. Experiments were conducted using helium and argon as driver and driven gases, respectively, at pressures ranging from 1.32 to 2.09 bar and temperatures from 1073 to 2126 K behind the reflected shock. High-speed imaging captured different diaphragm rupture profiles, classified into four distinct types based on their dynamics. Results indicate that the initial stages of diaphragm opening, including the rate and profile of opening, play crucial roles in resulting incident shock Mach number and test time. A sigmoid function was employed to fit the diaphragm opening profiles, allowing for accurate categorization and analysis. New correlations were developed to predict the incident shock attenuation rate and post-shock pressure rise, incorporating parameters such as diaphragm opening time, rupture profile constants, and normalized experimental Mach number. The results emphasize the importance of considering diaphragm rupture dynamics in shock tube experiments to achieve accurate predictions of shock parameters. 
\end{abstract}

\maketitle

\section{\label{sec:intro}INTRODUCTION}
Over the years, shock tubes have been employed to study the kinetics of high-temperature processes and to simulate high-speed flows over scaled models in ground test facilities \cite{Gaydon_1963,Kashif_2024,Janardhanraj_2022,Subburaj_2023}. They are designed to rapidly elevate the thermodynamic conditions of the test gas and maintain these conditions for a desired duration (called steady time or test time, $\Delta t$). This capability is vital for investigating rapid chemical reactions under 0-D conditions, thus eliminating transport process interference in chemical kinetics studies. Alternatively, shock tubes can generate high-enthalpy flows for aerodynamic research by expanding the high-pressure and high-temperature gas slug through a convergent-divergent nozzle. In its basic form, a conventional shock tube consists of a driver section filled with gas at high-pressure ($P_4$), separated from a driven section, filled with a test gas at low-pressure ($P_1$), by a diaphragm. When the diaphragm bursts due to the pressure difference, an incident shock wave propagates into the test gas at supersonic speed (represented in terms of Mach number as $M_1$), causing an almost immediate rise in temperature ($T_2$) and pressure ($P_2$) of the gas behind it. This shock wave is followed by a contact surface, which acts as an interface between the gases in the driver and driven sections. Upon reaching the driven section's end-wall, the shock wave reflects, further elevating the temperature and pressure near the end wall. Ideally, the thermodynamic conditions (temperature, $T_5$, and pressure, $P_5$) behind the reflected shock are constant and do not vary with location or time, and can be predicted using classical gas-dynamics theory. However, deviations from these ideal conditions, known as non-ideal effects, can significantly influence the system, leading to a reduction in shock strength as well as causing temporal and spatial variations in thermodynamic conditions. Several factors, such as the diaphragm opening process, viscous effects, and the three-dimensional nature of the flow have been attributed to the non-ideal effects \cite{Nativel_2020}.

Early studies addressing non-ideal effects in shock tubes focused on investigating the diaphragm material and the rupture initiation method to better understand the shock formation process in shock tubes. Campbell et al. \cite{Campbell_1965} measured the opening times of aluminum and copper diaphragms using flash photography and a photoelectric method, noting differences between natural bursting and pierce-assisted bursting. They suggested that pressurizing the driver section up to 90$\%$ of diaphragm rupture limit before piercing gives optimal performance. Rothkopf and Low \cite{Rothkopf_1974} described the diaphragm opening process for different materials used as diaphragms. They defined the shock formation distance as the distance from the diaphragm location where the shock speed reaches its maximum value. Glass et al. \cite{Glass_1955} showed that there are two main reasons for attenuation in a shock tube, namely, the shock formation process and the viscous effects due to the shock tube walls. Hickman et al. \cite{Hickman_1975} noted that thicker diaphragms, necessary for generating strong shocks, show considerable resistance to opening, impacting the rate at which the aperture opens. Kaneko et al. \cite{Kaneko_2016} investigated the diaphragm opening process in a miniature 10 mm internal diameter shock tube, discovering that faster opening rates of diaphragms lead to stronger shocks, whereas a smaller opened area leads to weaker shock waves. Wegener et al. \cite{Wegener_2000} and Mizoguchi et al. \cite{Mizoguchi_2006} explored the rupture mechanism of diaphragms in hypersonic impulse facilities and its effects on flow generation. Sasoh et al. \cite{Sasoh_1999} examined the effects of diaphragm rupture caused by an impinging projectile. Takahashi et al. \cite{Takahashi_2005} investigated shock tube operations where a diaphragm was ruptured by laser beam irradiation, with mylar and cellophone as diaphragm materials. While different methods of rupture initiation have been studied, the use of V-groove for metallic diaphragms and diaphragm cutters for non-metallic diaphragms (low-pressure operation) is commonplace.

White \cite{White_1958} measured the opening time of stainless steel diaphragms in a 3.25-inch square shock tube by tracking light transmission and proposed a model to account for the finite rupture time in large diameter shock tubes at high diaphragm pressure ratios, experimentally showing higher shock wave velocities than predicted by one-dimensional inviscid theory. Ikui et al. \cite{Ikui_1969, Ikui_1969_2} improved White's model with a multistage approach, relating shock formation distance ($x_f$) and hydraulic diameter ($D$) as $x_f.D \propto 0.88$. Rothkopf and Low \cite{Rothkopf_1976} qualitatively described diaphragm opening, agreeing with Drewry et al. \cite{Drewry_1965} on the diaphragm opening time ($T_{OP}$). Brun and Reboh \cite{Brun_1976} observed that longer diaphragm opening times increase shock formation distance, while Simpson et al. \cite{Simpson_1967} linked shock formation distance to shock speed and opening time, with Rothkopf et al. \cite{Rothkopf_1976} noting proportionality constants between 1 and 3. Ikui et al. \cite{Ikui_1979} found that shock waves become planar at one-fifth the shock formation distance. Rajagopal et al. \cite{Rajagopal_2012} simulated diaphragm openings in a micro shock tube, finding quadratic openings match experiments and reduce shock strength. Smith \cite{Smith_2017} used LS-DYNA to model diaphragm rupture in hypersonic tubes, and Lacey \cite{Lacey_2018} examined plastic yielding in thin diaphragms. Gaetani et al. \cite{Gaetani_2008} combined experimental and numerical methods, showing partial diaphragm openings diminish shock intensity and introduce pressure oscillations. Alves et al. \cite{Alves_2021} improved shock strength predictions using discharge-coefficient theory, while Houas et al. \cite{Houas_2012} found that reducing diaphragm area prolongs shock formation and reduces pressure jumps, sometimes producing compression waves instead of shocks.

More recently, Nativel et al. \cite{Nativel_2020} explored how non-ideal effects in shock tubes, such as the post-reflected-shock pressure rise (d$P^*$/d$t$) and attenuation rate ($AR$) of incident shock, depend on Mach numbers, pressures, and shock tube geometries. The study revealed a strong dependence on Mach number and a clear influence of shock tube geometry on d$P^*$/d$t$. It emphasized the importance of considering inherent facility effects for accurate shock tube combustion measurements, especially in the context of ignition delay times under various conditions. Fukushima et al. \cite{Fukushima_2020} focused on the effect of cellophane diaphragm opening processes in a short and low-pressure shock tube on shock wave formation. It was found that the self-shaping opening morphology of cellophane diaphragms significantly influences the early stages of shock wave formation. The study revealed that the initial stage of diaphragm opening, rather than the entire process, plays a dominant role in shock wave formation. This was deduced from high-speed visualization and analysis of the opening processes and the coalescence of compression waves. Kashif et al. \cite{Kashif_2024} studied the effect of diaphragm rupture pressure on the shock formation process in double-diaphragm shock tubes and compared the results with those obtained from single-diaphragm shock tubes. Subburaj et al. \cite{Janardhanraj_2021} observed that the formation of the shock wave is dominated by waves generated due to the finite rupture time of the diaphragm and their reflections from the walls of the shock tube. They proposed new correlations to predict the shock Mach number in both the shock formation and propagation regions for shock tubes of varying diameters.

From the existing studies in the literature, it is evident that the flow in a shock tube is significantly influenced by diaphragm opening effects during the initial stages of flow, known as the shock formation region, while boundary layer effects dominate the later stages of flow, referred to as the shock propagation region. Most of these studies either develop empirical relations between diaphragm rupture and shock formation distance or build correlations to predict shock attenuation due to wall effects, often neglecting the initial stages of shock formation. In the present work, the effects of diaphragm opening on shock parameters are investigated in detail. Various diaphragm opening patterns, including previously unstudied irregular ruptures, are captured to understand their influence on shock parameters near the driven section end-wall. The experiments employ helium as the driver gas and argon as the driven gas, conducted at pressures ranging from $P_5$ = 1.32 to 2.09 bar and temperatures from $T_5$ = 1073 to 2126 K. The diaphragm opening process is visualized using high-speed imaging, allowing for the categorization of different opening types based on their profiles. The influence of the diaphragm opening process on the incident shock Mach number, test time, and reflected shock pressure profile is analyzed. A sigmoid function is employed to fit and quantify the diaphragm opening profiles accurately. Finally, correlations are developed to enhance predictions of $AR$ and d$P^*$/d$t$ using initial operating conditions, shock wave parameters, and the diaphragm opening profiles.

\begin{figure*}
    \centering
    \includegraphics[width=0.9\textwidth]{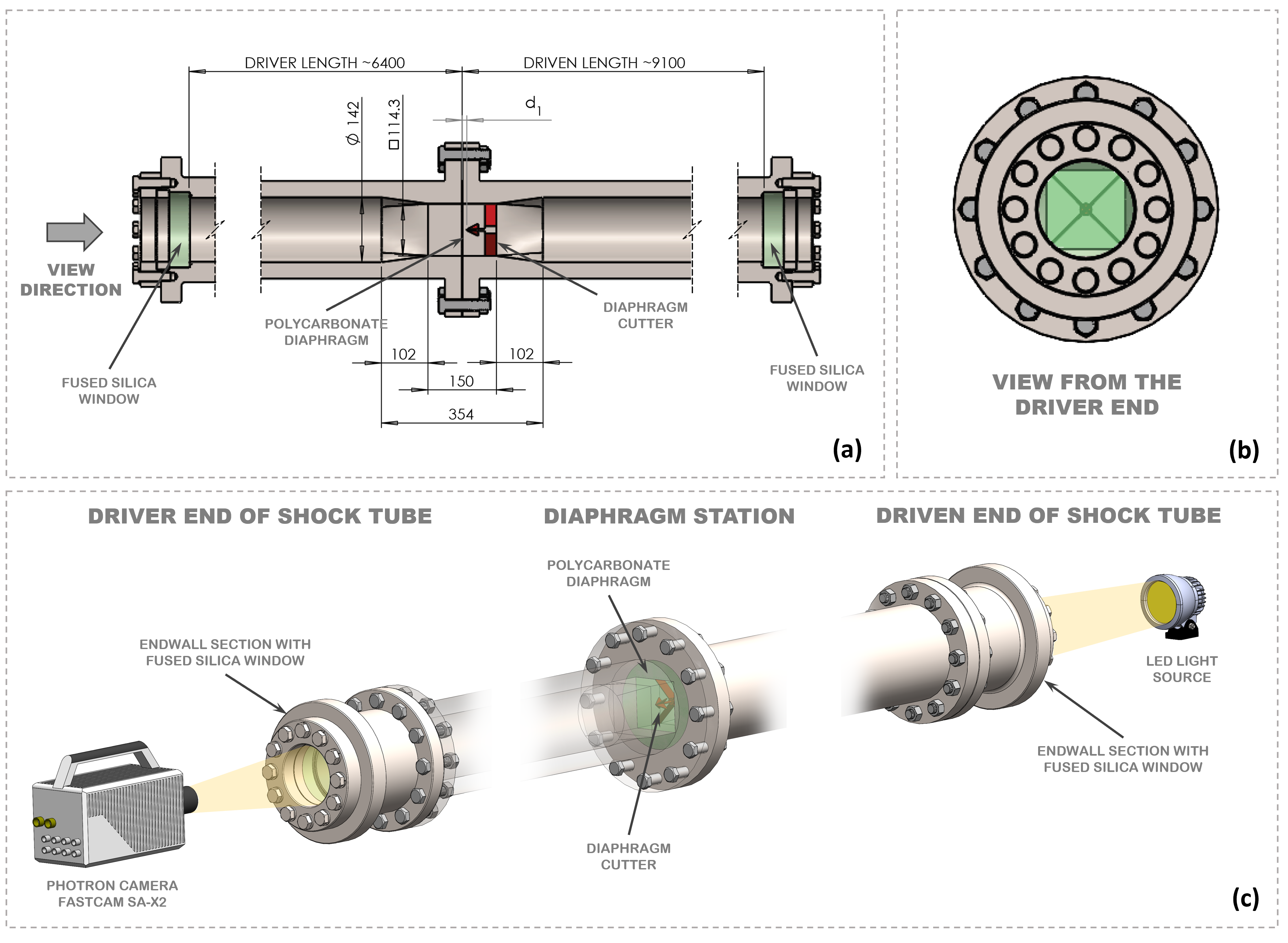}
    \caption{Schematic diagrams of the LPST and the high-speed imaging setup. (a) Cross-sectional view of LPST showing the diaphragm station and the transparent end walls. All dimensions are in millimeters. (b) View of the diaphragm section from the driver end of the shock tube. A small portion of the diaphragm at the four corners is not visible because of the square to circle transition. (c) Schematic diagram showing the imaging setup with the high-speed camera, light source, and location of the diaphragm. Note: The light is represented using a yellow color to enhance visibility in the schematic, despite being generated by a white LED.}
    \label{fig:exp_setup}
\end{figure*}

\section{\label{sec:methods}METHODOLOGY}
\subsection{\label{sec:LPST}Shock tube imaging setup}
Experiments were performed in the Low-Pressure Shock Tube (LPST) facility at KAUST. A schematic diagram of the shock tube and imaging setup  is shown in Figure \ref{fig:exp_setup}. The driver and driven sections of this facility are each 9.1 m long and have an inner diameter of 14.2 cm. The modular design of the driver section facilitates variation in the driver length to obtain the desired test times in the experiments. For the present work, the length of the driver section was maintained at 6.4 m (limited by the space requirement for the end wall imaging setup on the driver side) to obtain a maximum test time of 4.95 milliseconds. The diaphragm station of the shock tube, as shown in Figure \ref{fig:exp_setup}a, is 354 mm in length. It transitions from a circular cross-section in the driver section (diameter of 14.2 cm) to a square cross-section (side of 11.4 cm) and back to the circular cross-section in the driven section. A diaphragm cutter is employed to puncture the polycarbonate diaphragms used for shock generation. The square cross-section around the diaphragm location is intended to facilitate a uniform diaphragm rupture and petal formation. For a previous work, a circular optical section was fabricated that can be mounted at the end of the driven section of the shock tube \cite{Labastida_2018}. The same optical section was used for the present work, and another similar optical section was fabricated to mount at the end of the driver section. The optical test sections are made of SS316 and are designed for a maximum pressure of 30 bar. Optical quality fused silica windows with a thickness of 45.2 mm were used as the end wall on both the driver and driven sections. The optical sections were designed to enable visualization of the entire circular section of the shock tube. A stepped configuration was incorporated so that the inner surface of the end wall aligns with the edge of the slit window. 

\begin{table*}
\centering
\begin{ruledtabular}
\begin{tabular}{ccccccccc}
Test identifier & $P_4$, psi & $P_1$, Torr & $M_1$, - & $P_2$, bar & $T_2$, K & $P_5$, bar & $T_5$, K & Test time, ms \\ \hline
A01 & 16.5 & 65 & 2.02 & 0.42 & 646 & 1.32 & 1073 & 4.949 \\
A02 & 17.0 & 65 & 2.13 & 0.47 & 692 & 1.56 & 1181 & 4.818 \\
A03 & 17.8 & 65 & 2.15 & 0.48 & 702 & 1.62 & 1205 & 4.846 \\
A04 & 18.2 & 65 & 2.15 & 0.48 & 703 & 1.62 & 1207 & 4.740 \\
A05 & 18.2 & 65 & 2.17 & 0.49 & 710 & 1.66 & 1224 & 4.796 \\
A06 & 18.5 & 65 & 2.18 & 0.49 & 716 & 1.69 & 1237 & 4.662 \\
A07 & 27.4 & 65 & 2.22 & 0.51 & 733 & 1.79 & 1280 & 4.463 \\
A08 & 29.0 & 65 & 2.29 & 0.55 & 762 & 1.95 & 1349 & 4.181 \\
A09 & 29.2 & 65 & 2.33 & 0.56 & 780 & 2.04 & 1390 & 4.060 \\ \hline
A10 & 24.2 & 35 & 2.34 & 0.31 & 784 & 1.11 & 1401 & 2.964 \\
A11 & 16.9 & 35 & 2.48 & 0.35 & 854 & 1.33 & 1566 & 3.543 \\
A12 & 16.5 & 35 & 2.49 & 0.35 & 854 & 1.33 & 1566 & 3.412 \\
A13 & 17.1 & 35 & 2.50 & 0.35 & 864 & 1.36 & 1590 & 3.511 \\
A14 & 17.4 & 35 & 2.52 & 0.36 & 870 & 1.38 & 1605 & 3.546 \\
A15 & 17.5 & 35 & 2.52 & 0.36 & 871 & 1.38 & 1608 & 3.435 \\
A16 & 26.3 & 35 & 2.56 & 0.37 & 892 & 1.45 & 1657 & 3.353 \\
A17 & 35.3 & 35 & 2.73 & 0.42 & 979 & 1.73 & 1866 & 2.941 \\
A18 & 35.4 & 35 & 2.74 & 0.43 & 987 & 1.75 & 1883 & 2.726 \\
A19 & 27.5 & 35 & 2.80 & 0.44 & 1014 & 1.84 & 1949 & 2.883 \\
A20 & 40.0 & 35 & 2.80 & 0.45 & 1019 & 1.86 & 1960 & 2.523 \\
A21 & 30.0 & 35 & 2.87 & 0.47 & 1057 & 1.98 & 2051 & 2.643 \\
A22 & 30.4 & 35 & 2.88 & 0.47 & 1060 & 1.99 & 2057 & 2.618 \\
A23 & 30.8 & 35 & 2.88 & 0.47 & 1063 & 2.00 & 2067 & 2.613 \\
A24 & 32.4 & 35 & 2.93 & 0.49 & 1088 & 2.09 & 2126 & 2.483 \\          
\end{tabular}
\end{ruledtabular}
\caption{The driver and driven pressures used in the experiments and the measured incident shock Mach number in the experiments. The shock parameters estimated from the incident shock Mach number are also indicated in the table. The experiments are grouped based on the driven gas pressures.}
\label{tab:exp_Cond}
\end{table*}

\subsection{\label{sec:Oper_cond}Operating conditions and measurements}
The diaphragm opening time, opening profile, and driver rupture pressure vary based on the thickness of the polycarbonate diaphragm and the initial distance between the diaphragm and the cutter. To initiate the rupture of the diaphragm, a small distance (represented by $d_1$ in Figure \ref{fig:exp_setup}a) is initially maintained between the tip of the cutter and the diaphragm to allow space for diaphragm bulge. As the gas fills the driver section, the diaphragm bulges towards the driven section and is pierced by the cutter, initiating rupture. In the present study, the thicknesses of the polycarbonate diaphragms ranged from 0.005 to 0.015 inches, while the distance $d_1$ was varied between 7 and 10 mm to obtain different burst conditions. The driver pressure was recorded using a pressure sensor, and high-speed images of the diaphragm opening process were captured in each experiment. Table \ref{tab:exp_Cond} shows the values of driver pressure ($P_4$) and driven pressure ($P_1$) for the experiments reported here. In all experiments, helium was used as the driver gas and argon as the driven gas.

The incident shock speed was measured by the time-of-flight method using a series of five piezoelectric pressure transducers (Model 113B26, PCB Piezotronics, USA) mounted over the last 1.5 meters of the driven section of the shock tube. The attenuation of the incident shock wave was estimated for each experiment, and the shock velocity at the end wall was obtained through linear extrapolation, with velocity errors of less than 0.2\%. The attenuation of the shock is represented as a percentage per meter (\%/m). Thermodynamic conditions behind the incident and reflected shock were estimated from the incident shock speed, thermodynamic parameters of the shock-heated gas (Ar), and shock jump relations using a MATLAB code. The post-reflected-shock pressure rise (d$P^*$/d$t$) was measured following procedures used in previous studies \cite{Subburaj_2023} with a linear fit over the test time duration after the reflected shock wave. The post-shock pressure rise (d$P^*$/d$t$) is represented as a percentage per millisecond (\%/ms). 

\subsection{\label{sec:imag_proc}Image acquisition and processing}
The view of the diaphragm station from the driver end of the shock tube is shown in Figure \ref{fig:exp_setup}b. A small portion of the diaphragm at the four corners of the square is not visible in the end view due to the transition from square to circular cross-section. Figure \ref{fig:exp_setup}c displays the setup for imaging the diaphragm opening process during the shock tube experiments. A 5-watt white LED light source (Model: XLamp XML series, CreeLED Inc., USA) was positioned on the driven end for illumination, while a high-speed camera was placed on the driver side, as shown in Figure \ref{fig:exp_setup}b. This arrangement allows for complete longitudinal visualization of the diaphragm opening process in the square cross-section. A Photron FASTCAM SA-X2 camera was used to capture images at 81,000 frames per second, with the pressure rise from the incident shock wave serving as the trigger to record the camera signals. The images were cleaned and enhanced using a method similar to that described in a previous work \cite{Janardhanraj_2021}. An edge detection algorithm based on the Sobel filter was employed to track the diaphragm edges post-rupture in each image frame. In some cases, the threshold levels of the filter had to be adjusted manually to detect the edge due to the background noise. A polygon fit was used to estimate the opened aperture area of the diaphragm from the detected edges. 

\begin{figure*}
    \centering
    \includegraphics[width=0.9\textwidth]{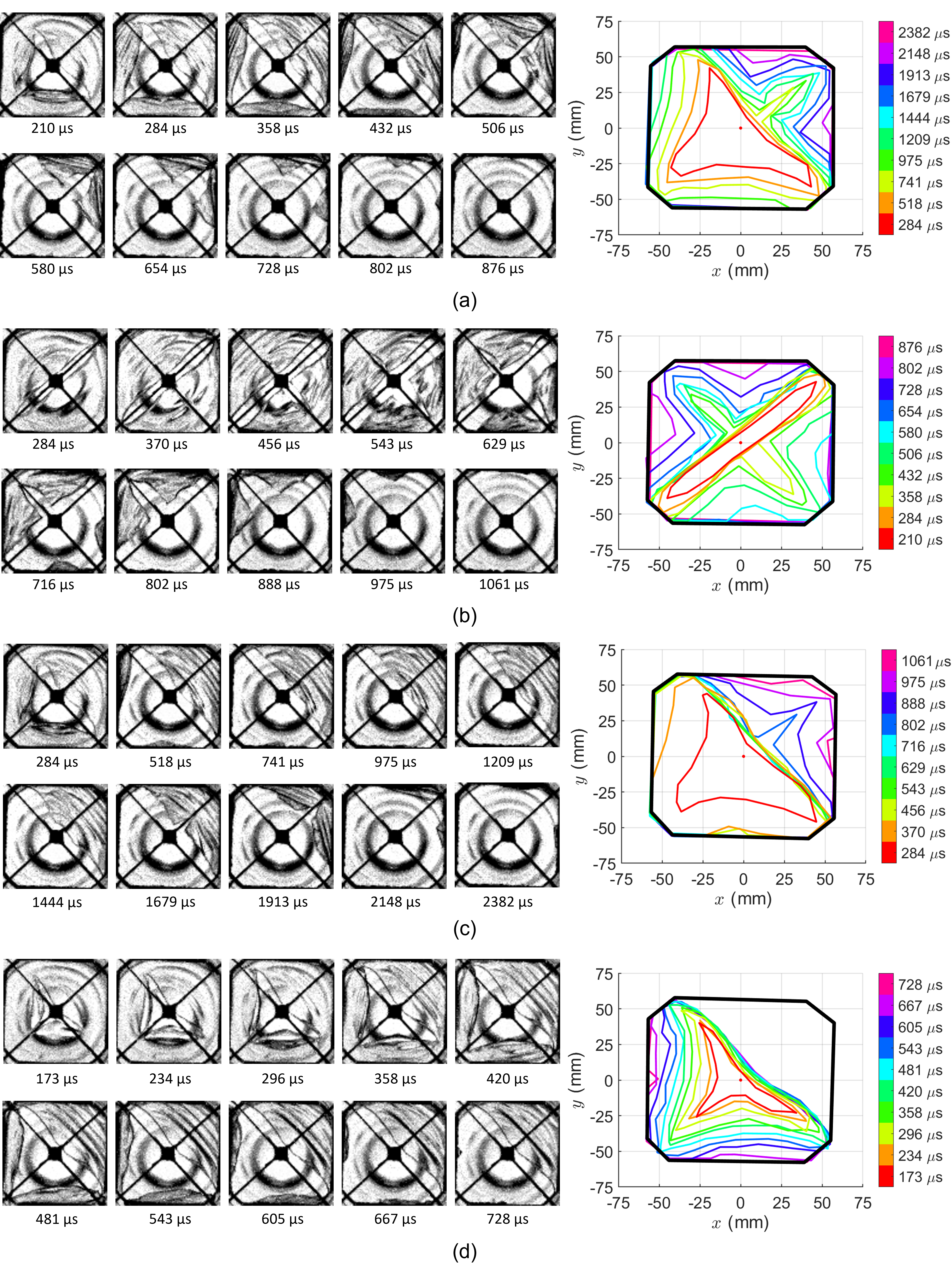}
    \caption{Different diaphragm opening profiles observed in the experiments. On the left, images captured using the high-speed camera at different time instants are shown. The opened aperture captured based on the edge detection algorithm is shown on the right. (a) Type-1 opening profile (b) Type-2 opening profile (c) Type-3 opening profile (d) Type-4 opening profile}
    \label{fig:open_profile}
\end{figure*}

\begin{figure*}
    \centering
    \includegraphics[width=0.75\textwidth]{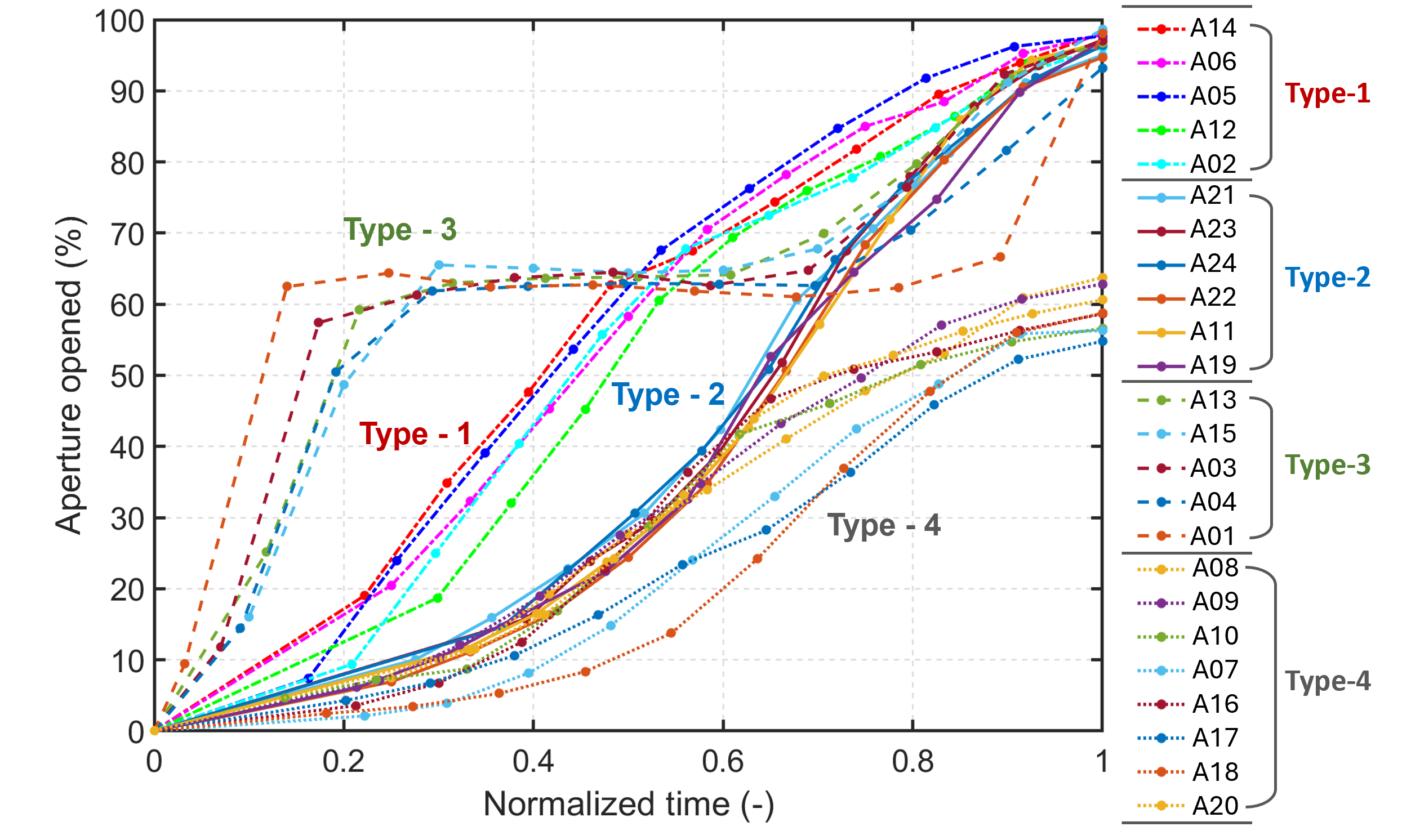}
    \caption{Plot showing the percentage of opening as a function of normalized time. Time is normalized with opening time of the diaphragm  $T_{op}$. The tests are grouped based by opening type and the number in the gray circle next to the curves indicates the corresponding type of opening.}
    \label{fig:aper_time}
\end{figure*}

\section{\label{sec:results}RESULTS}
\subsection{\label{sec:postproc}Types of diaphragm opening}
Figure \ref{fig:open_profile} depicts the diaphragm opening process observed in the experiments. Adjacent to the images obtained from the high-speed camera, there are plots showing the polygons derived from edge detection for each image, illustrating the evolving aperture post-rupture. The first frame in which the tear in the diaphragm appears is given a timestamp $t=0$. The experiments revealed four distinct rupture profiles, as displayed in the subplots of Figure \ref{fig:open_profile}. Type-1 rupture, shown in Figure \ref{fig:open_profile}a, exhibits a relatively uniform opening profile, though it is asymmetrically skewed toward one half of the cross-section. Figure \ref{fig:open_profile}b presents the type-2 rupture, characterized by a more longitudinal opening compared to type-1. Initially, the rupture is almost symmetric about the cross-section's diagonal, but as time progresses, one half opens more rapidly than the other. Type-3 rupture, shown in Figure \ref{fig:open_profile}c, indicates a delayed rupture, with only one half of the cross-section opening initially, followed by a temporary pause before the other half of the diaphragm opens. Lastly, Figure \ref{fig:open_profile}d illustrates the type-4 rupture, where only one half of the cross-section opens, while the remaining half obstructs the flow.

The diaphragm rupture dynamics can also be analyzed graphically by tracking the aperture opened over time to complement the qualitative observations presented in Figure \ref{fig:open_profile}. Figure \ref{fig:aper_time} provides a detailed depiction of the aperture opening as a function of time, enabling a clear distinction among the four identified types of diaphragm ruptures. The vertical axis of the plot is normalized against the total cross-sectional area at the diaphragm station. To facilitate comparison, the time axis is normalized with the duration required for each diaphragm to transition from the initial breach, $t=0$, to its completely opened state (also called the diaphragm opening time, $T_{op}$). This normalization of time allows for an intuitive comparison of the opening dynamics across different diaphragm types. Figure \ref{fig:aper_time} illustrates that the type-1 rupture exhibits a rapid increase in the opened aperture area, indicating a faster spreading of the rupture across the diaphragm's surface. In contrast, the type-2 opening demonstrates a less steep ascent compared to the type-1 rupture, suggesting a more gradual opening process. Furthermore, the distinctive characteristic of the type-3 opening, a pronounced delay in the opening sequence, is captured effectively in the plot. This manifests as a plateau in the curve, signifying a period during which the opening progression halts before resuming to complete the rupture process. A type-4 rupture is characterized by a partially opened diaphragm, with the aperture reaching slightly more than half of its full opening. This graphical analysis enables a quantitative assessment and comparison of the dynamic behavior of diaphragm openings, providing deeper insights into the previously outlined classifications.

\begin{figure*}
    \centering
    \includegraphics[width=\textwidth]{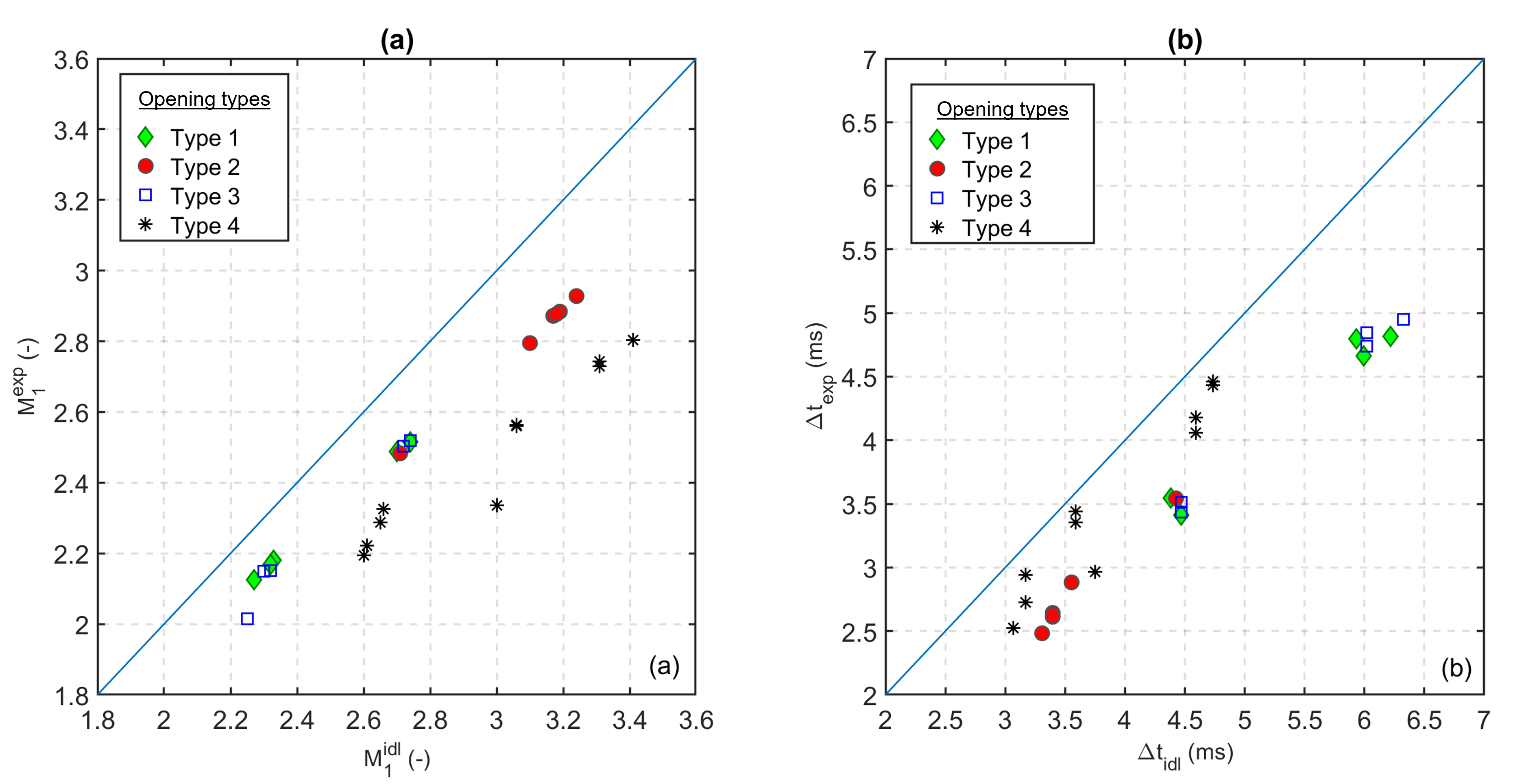}
    \caption{Comparison of experimentally obtained Mach number and test time with those obtained using 1-D inviscid shock relations (ideal theory). The solid line is an identify line using the ideal theory. (a) Comparison between ideal and experimental shock Mach number. (b) Comparison between ideal and experimentally obtained test time.}
    \label{fig:idl_comp}
\end{figure*}

\begin{figure*}
    \centering
    \includegraphics[width=0.9\textwidth]{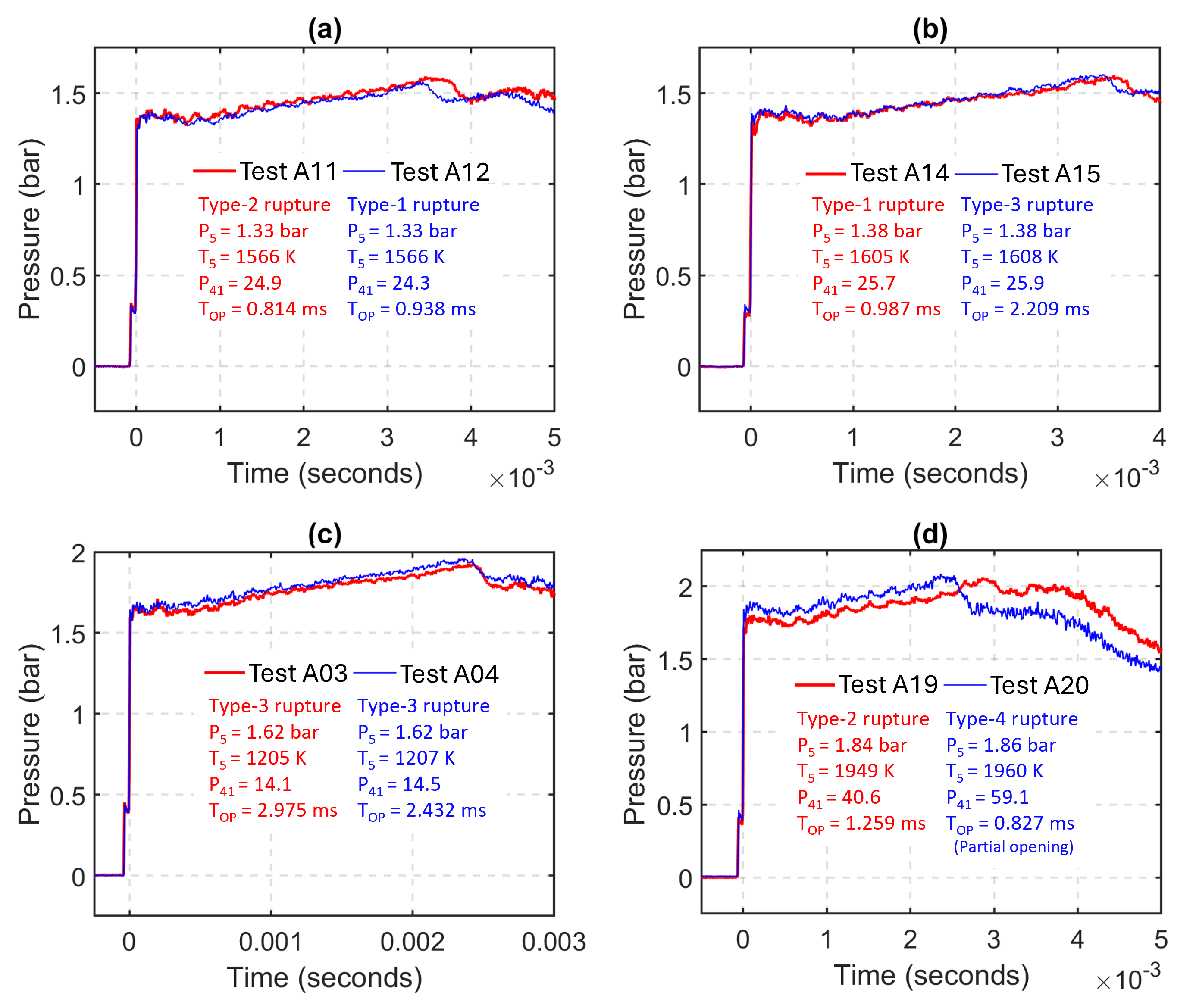}
    \caption{Plots showing comparison between the tests with similar incident shock Mach number but different rupture profiles. (a) Comparing type-1 and type-2 rupture cases. (b) Comparing type-1 and type-3 rupture cases. (b) Comparing two type-3 rupture cases with different diaphragm opening times. (d) Comparing type-2 and type-4 ruptures cases.}
    \label{fig:P5}
\end{figure*}

\subsection{\label{sec:openingpara}Influence of diaphragm opening process on $M_1$ and $\Delta t$}
Table \ref{tab:exp_Cond} shows the incident shock Mach numbers ($M_1$) and test times ($\Delta t$) measured in the experiments. For every test condition, using the initial driver and driven pressures, the length of the shock tube sections, and the gases used in the driver and driven sections, the ideal incident shock Mach number ($M_1^{idl}$) and ideal test time ($\Delta t_{idl}$) were computed using a WENO shock tube code. The code assumes a one-dimensional inviscid flow with an instantaneous diaphragm rupture. Additional details of the numerical code are described in a previous work \cite{Kashif_2024}. Figure \ref{fig:idl_comp} compares the experimental data with the ideal values of incident shock Mach number and test time for the various diaphragm opening types. As depicted in Figure \ref{fig:idl_comp}a, the experimental $M_1$ values for type-1, type-2, and type-3 openings align more closely with their ideal predictions compared to type-4 openings. This shows that incomplete rupture of the diaphragm can result in a significant drop in the shock Mach number obtained in the experiments. The deviation between experimental and ideal Mach numbers also tend to increase with the shock Mach number. A different pattern emerges in the comparison of test times shown in Figure \ref{fig:idl_comp}b. Type-4 ruptures, which are deemed non-ideal, displayed test times that are closer to the ideal predictions than those of type-1, type-2, and type-3 openings. This phenomenon stems from a delayed interaction between the reflected shock wave and the contact surface in the type-4 ruptures. Although a slower-moving incident shock is seen in type-4 ruptures, a partially opened diaphragm can significantly restrict the driver gas mass flow, reducing the speed of the contact surface. 

\begin{figure}
    \centering
    \includegraphics[width=\columnwidth]{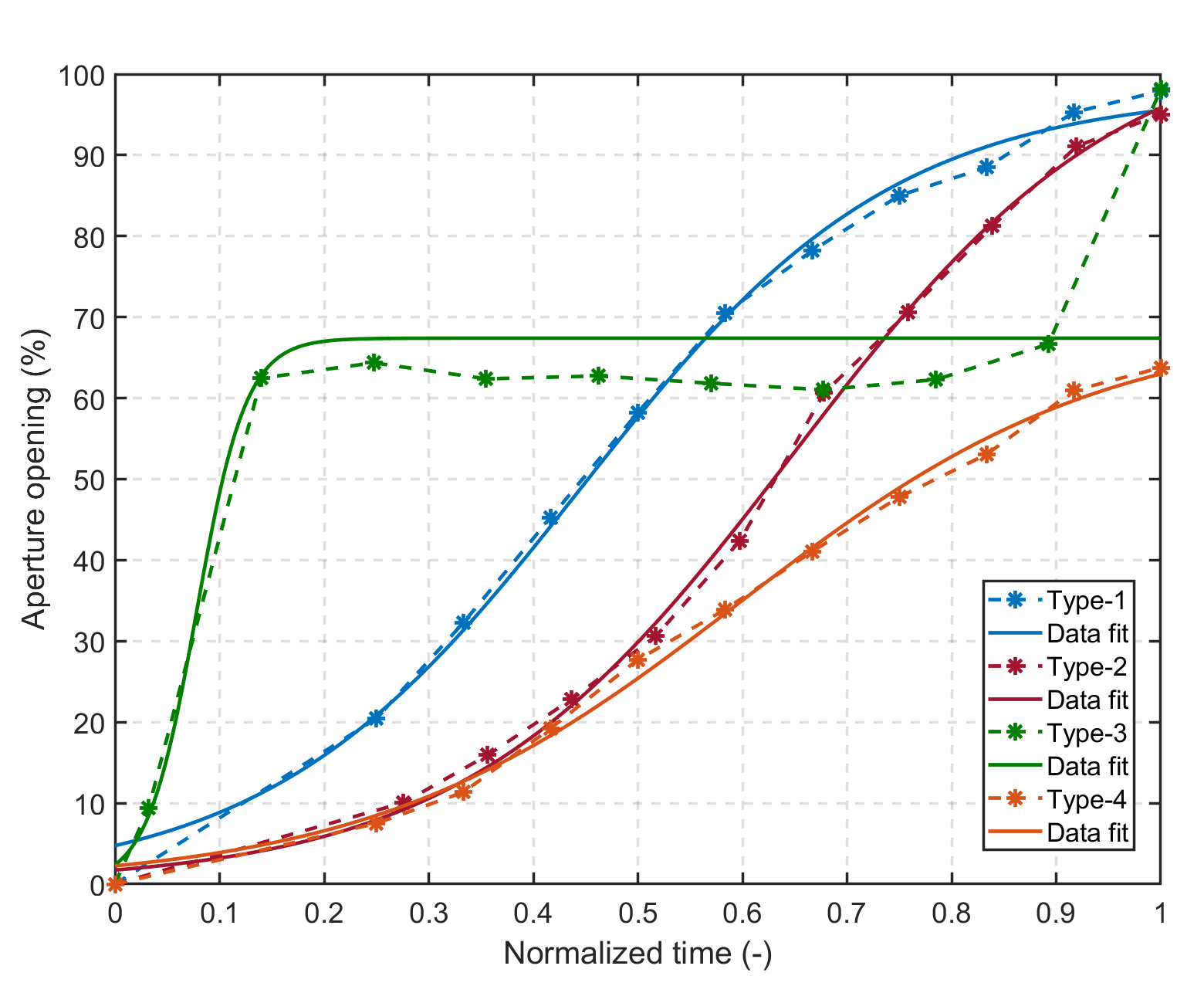}
    \caption{Plot showing curve fitting using the sigmoid function (Eq. \ref{eqn:sigmoid}) for the four types of diaphragm opening profiles.}
    \label{fig:sigmoid}
\end{figure}

\subsection{\label{sec:P5region}Influence of diaphragm opening process on $P_5$ profile}
To understand the influence of the diaphragm opening process on the pressure profile behind the reflected shock, test cases with similar incident shock Mach numbers but different rupture profiles were considered. Due to the complexity and variability in the diaphragm rupture process, achieving precise control over the shock Mach number is challenging. Consequently, only a limited number of comparisons could be made. Figure \ref{fig:P5} illustrates four different comparisons of reflected shock pressure profiles corresponding to various diaphragm rupture cases. Figure \ref{fig:P5}a compares test cases with type-1 and type-2 ruptures. Test A11, characterized by a slightly higher diaphragm rupture pressure, exhibited a shorter diaphragm opening time compared to Test A12. But the rate of opening differed significantly between the two tests. Type-1 ruptures demonstrated a faster initial opening rate compared to type-2 ruptures. For instance, to achieve 50\% aperture opening, Test A12 (type-1 rupture) required approximately 0.47 ms ($\approx 0.5 \times T_{OP}$), whereas Test A11 required about 0.57 ms ($\approx 0.7 \times T_{OP}$). This indicates that not only the diaphragm opening time but also the rate of opening is a crucial parameter influencing the shock dynamics.

Figure \ref{fig:P5}b compares Test A14 and Test A15, which display type-1 and type-3 ruptures, respectively. Despite the substantial difference in diaphragm opening times (0.987 ms for Test A14 and 2.209 ms for Test A15) the post-shock conditions were remarkably similar. Examining the initial stages of diaphragm rupture, 50\% of aperture opening was reached in approximately 0.40 ms ($\approx 0.4 \times T_{OP}$) in Test A14 and about 0.44 ms ($\approx 0.2 \times T_{OP}$) in Test A15, followed by a stagnation period. This highlights the importance of the initial stages of diaphragm opening in determining the overall rupture dynamics. Further emphasizing this point, Figure \ref{fig:P5}c presents two type-3 ruptures observed in Tests A03 and A04. Despite a difference of about half a millisecond in diaphragm opening times, the initial opening profile up to 60\% opening was similar in both cases, occurring between 0.75-0.80 ms. This similarity in the early stages of opening underscores the significant impact of initial rupture on the shock dynamics.

Finally, Figure \ref{fig:P5}d compares Tests A19 and A20, illustrating the difference between shock tube flow past a partially opened diaphragm and a fully opened diaphragm. Despite having very similar incident shock Mach numbers, the diaphragm rupture pressures varied significantly. Test A20, characterized by partial opening, exhibited a higher $P_{41}$ compared to Test A19. This comparison underscores that the extent of diaphragm opening profoundly influences the flow dynamics within the shock tube. Overall, it is evident that the initial stages of diaphragm rupture, the rate and profile of opening, and the final percentage of diaphragm opening all play pivotal roles in shaping the flow characteristics within the shock tube. These factors collectively influence the shock wave behavior, the reflected shock pressure profile, and ultimately the performance and outcomes of shock tube experiments.

\begin{table*}
\centering
\begin{ruledtabular}
\begin{tabular}{ccccccccccc}
Test No. & Rupture type & $P_1$ & $M_1^{exp}$ & $M_1^{idl}$ & $AR$ & d$P^*/$d$t$ & $T_{op}$ & \multicolumn{3}{c}{Opening profile fit parameters} \\
 &  & Torr & - &  - & \% /$m$ & \% /$ms$ & $ms$ & $c_1$ & $c_2$ & $c_3$ \\ \hline
A14 & 1 & 35 & 2.52 & 2.74 & 2.508 & 5.067 & 0.987 & 96.678 & 6.277 & 0.417 \\
A06 & 1 & 65 & 2.18 & 2.33 & 1.973 & 3.542 & 0.876 & 97.894 & 6.675 & 0.446 \\
A05 & 1 & 65 & 2.17 & 2.32 & 2.017 & 3.479 & 1.049 & 97.183 & 7.310 & 0.422 \\
A12 & 1 & 35 & 2.49 & 2.70 & 2.192 & 4.564 & 0.938 & 95.749 & 7.057 & 0.477 \\
A02 & 1 & 65 & 2.13 & 2.27 & 2.190 & 1.526 & 1.111 & 92.411 & 7.365 & 0.435 \\ \hline
A21 & 2 & 35 & 2.87 & 3.17 & 2.413 & 6.854 & 1.061 & 106.241 & 6.319 & 0.649 \\
A23 & 2 & 35 & 2.88 & 3.19 & 2.036 & 6.906 & 0.901 & 108.627 & 6.830 & 0.666 \\
A24 & 2 & 35 & 2.93 & 3.24 & 2.407 & 6.480 & 0.864 & 106.784 & 6.485 & 0.652 \\
A22 & 2 & 35 & 2.88 & 3.18 & 2.313 & 2.883 & 1.024 & 106.658 & 6.847 & 0.676 \\
A11 & 2 & 35 & 2.48 & 2.71 & 2.233 & 2.389 & 0.814 & 113.071 & 6.463 & 0.692 \\
A19 & 2 & 35 & 2.80 & 3.10 & 2.081 & 2.974 & 1.259 & 111.638 & 6.094 & 0.690 \\ \hline
A13 & 3 & 35 & 2.50 & 2.72 & 2.042 & 4.967 & 2.382 & 76.078 & 14.467 & 0.164 \\
A15 & 3 & 35 & 2.52 & 2.74 & 1.830 & 5.090 & 2.209 & 76.366 & 14.886 & 0.179 \\
A03 & 3 & 65 & 2.15 & 2.30 & 1.837 & 3.996 & 2.975 & 73.293 & 25.711 & 0.131 \\
A04 & 3 & 65 & 2.15 & 2.32 & 1.680 & 2.041 & 2.432 & 70.619 & 19.858 & 0.155 \\
A01 & 3 & 65 & 2.02 & 2.25 & 1.253 & 1.954 & 4.580 & 67.402 & 42.289 & 0.078 \\ \hline
A08 & 4 & 65 & 2.29 & 2.65 & 2.153 & 3.399 & 0.728 & 69.331 & 5.687 & 0.596 \\
A09 & 4 & 65 & 2.33 & 2.66 & 2.070 & 3.640 & 0.716 & 66.826 & 6.290 & 0.561 \\
A10 & 4 & 35 & 2.34 & 3.00 & 2.159 & 1.740 & 1.148 & 57.259 & 8.052 & 0.520 \\
A07 & 4 & 65 & 2.22 & 2.61 & 2.527 & 1.560 & 0.987 & 59.826 & 7.906 & 0.625 \\
A16 & 4 & 35 & 2.56 & 3.06 & 2.408 & 2.544 & 0.975 & 57.549 & 9.793 & 0.513 \\
A17 & 4 & 35 & 2.73 & 3.31 & 2.194 & 2.557 & 0.962 & 64.917 & 5.541 & 0.676 \\
A18 & 4 & 35 & 2.74 & 3.31 & 2.269 & 2.423 & 1.074 & 64.803 & 8.088 & 0.695 \\
A20 & 4 & 35 & 2.80 & 3.41 & 2.439 & 3.186 & 0.827 & 61.113 & 8.157 & 0.532 \\       
\end{tabular}
\end{ruledtabular}
\caption{Table showing the values of the shock parameters, incident shock attenuation rate ($AR$), post-shock pressure rise, diaphragm opening times, and opening profile fit parameters for each experiment. The tests are grouped based on the rupture type.}
\label{tab:estimates}
\end{table*}

\section{\label{sec:disc}DISCUSSION}

\subsection{\label{sec:postshock}Sigmoidal fit for the diaphragm opening profiles}
Campbell et al. \cite{Campbell_1965} proposed an empirical relation to fit the diaphragm opening profiles in shock tubes. This relation featured an S-shaped curve, which was derived based on several assumptions. These assumptions included: (a) The shock tube cross-section at the diaphragm station is considered to be square with an area equivalent to the circular cross-section area, providing a simplified model for analysis. (b) The diaphragm is assumed to rupture instantaneously without any prior deformation, which disregards any gradual weakening or pre-rupture bulging of the diaphragm material. (c) Upon rupture, the diaphragm is believed to form four identical triangular petals, which simplifies the complexity of the actual rupture dynamics. (d) The force acting on each petal is assumed to be a linear function of the opening area, starting from a maximum initial value and decreasing to zero when the diaphragm is fully open. At any given moment, the force is considered to be uniformly distributed over the petal surface, acting at its centroid. (e) The moment due to bending stresses in the petal is assumed to be constant during the rupture process. This simplified assumption neglects the variability in stress distribution that can occur during the dynamic rupture.

Based on these assumptions, the motion of each diaphragm petal can be represented by the following equation:

\begin{equation}
\frac{d^2\theta}{dt^2} = \frac{4P_4 \cos\theta}{\rho b t}
\label{eqn:campbell}
\end{equation}

where $P_4$ is the diaphragm rupture pressure, $\rho$ is the density of the material, $b$ is the base width of the diaphragm petal, and $t$ is the thickness of the diaphragm. This relation generates a single type of curve for given material properties and does not account for the irregular ruptures often encountered in practical scenarios. Therefore, it is more useful to fit a generalized S-shaped curve to the diaphragm opening profile in each experiment. To address this need, an appropriate fitting model based on the sigmoid function is employed here to characterize the diaphragm opening profiles accurately.

A sigmoid function is defined as a mathematical function that yields an S-shaped, or sigmoid, curve. It is a bounded, differentiable, and real function defined for all real input values, possessing a non-negative derivative at each point, and exactly one inflection point. The general form of the sigmoid function can be expressed as:

\begin{equation}
y = \frac{c_1}{1+e^{-c_2(x-c_3)}}
\label{eqn:sigmoid}
\end{equation}

In this equation, $c_1$, $c_2$, and $c_3$ are constants adjusted to alter the shape of the S-curve. When applied to the diaphragm opening profile, these constants can be correlated to various aspects of the diaphragm opening process. The constant $c_1$ represents the maximum aperture opening, with a higher value corresponding to a larger final opening of the diaphragm. The constant $c_2$ influences the sharpness of the curve's transition, representing the stretch of the S-curve. Higher values of $c_2$ indicate faster transitions, which is particularly noticeable in type-3 ruptures where the opening occurs more rapidly. The constant $c_3$ determines the inflection point or the symmetry point of the curve. A higher value of $c_3$ suggests that the initial rupture phase is slower compared to the latter stages, indicating a delayed but accelerated final opening.

Figure \ref{fig:sigmoid} demonstrates the application of the sigmoid function to approximate the initial diaphragm opening profiles for the four types of diaphragm ruptures observed in experiments. The curve fits for type-1, type-2, and type-4 ruptures align closely with the empirical data, accurately reflecting the dynamics of these rupture types. The approximation for type-3 ruptures accurately represents the early stages of opening and the pause phase but is less accurate for the latter stages of opening. Prior research, including findings by Fukushima et al. \cite{Fukushima_2020}, has indicated that the initial phase of diaphragm rupture predominantly influences shock wave formation. Therefore, a data fit that effectively captures the initial phase of diaphragm rupture is considered sufficient for characterizing the type-4 profile.

Table \ref{tab:estimates} lists the fitting parameters derived for the various test cases, corresponding to the aperture opening curves displayed in Figure \ref{fig:aper_time}. These parameters provide a quantitative basis for comparing the different rupture types and understanding the influence of diaphragm dynamics on shock tube performance. By employing the sigmoid function as a fitting model, a more accurate and generalized fit for diaphragm rupture can be achieved. This model also provides qualitative prediction of the diaphragm rupture which can be utilized to enhance the predictions of the incident shock wave attenuation and post-shock pressure rise, as described in the subsequent sections.

\begin{figure*}
    \centering
    \includegraphics[width=\textwidth]{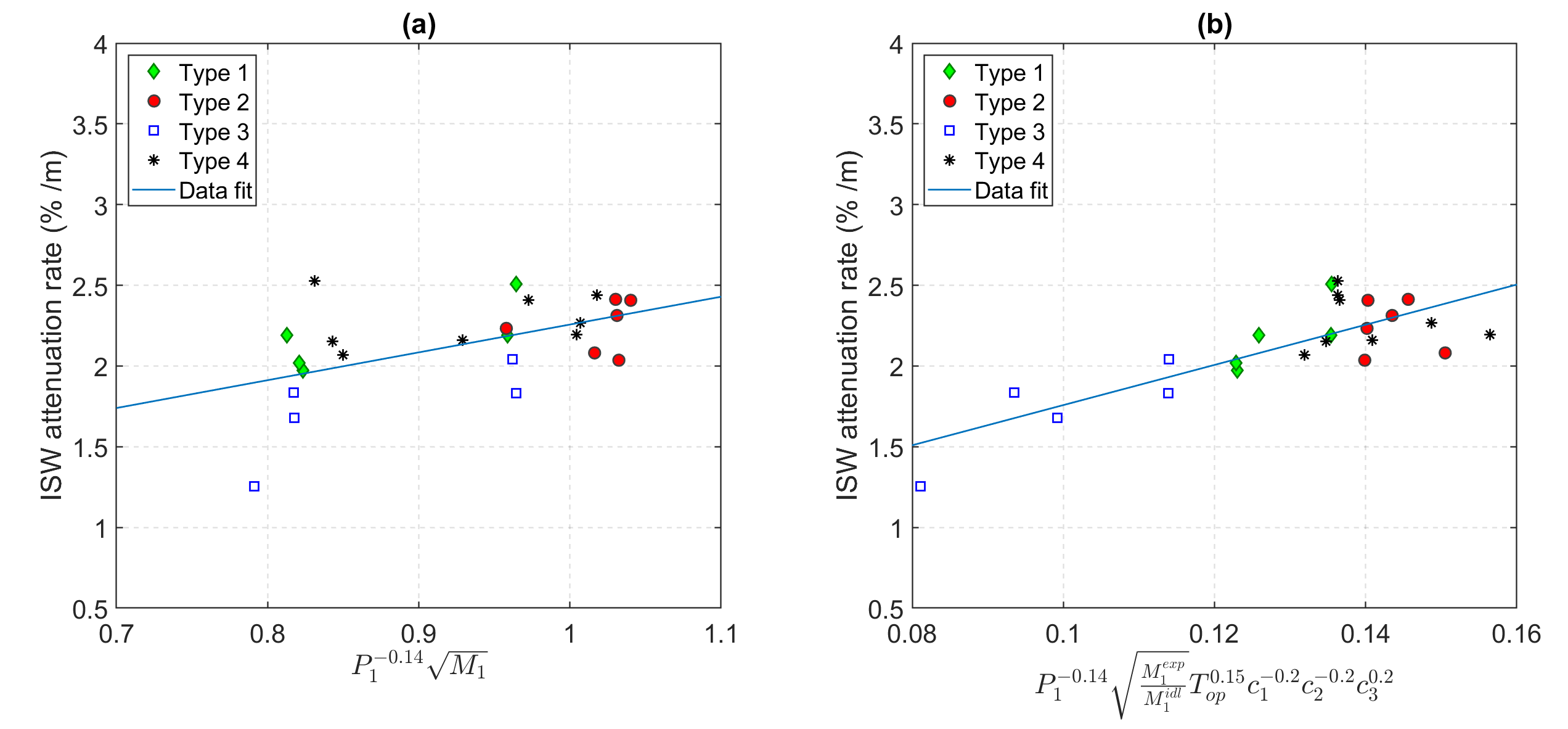}
    \caption{Comparison between existing correlation for incident shock wave (ISW) attenuation rate and the newly developed correlation. (a) Incident-shock attenuation data as a function of $P_1$ and $M_1$ based on previous work\cite{Nativel_2020}. (b) Incident-shock attenuation data as a function of shock parameters, diaphragm opening time, and opening profile parameters.}
    \label{fig:ISW_correlation}
\end{figure*}

\subsection{\label{sec:atten}Correlation for incident shock attenuation rate}
Correlations for incident shock attenuation have historically been formulated based primarily on viscous effects. According to the attenuation theory proposed by Mirels \cite{Mirels_1963, Mirels_1964}, attenuation resulting solely from viscous effects should vary with \( P_1 \) and \( M_1 \). Petersen \& Hanson \cite{Petersen_2001} explained that the exponent for \( P_1 \) dependence (i.e., 0.14) originates from a friction correlation. He further described the Mach number dependence as a result of boundary-layer growth. As the boundary layer expands, the wall surface temperature increases, becoming the dominant factor for attenuation, scaling with the square root of the Reynolds number. Consequently, it was demonstrated that incident shock wave attenuation rate ($AR$) primarily depends on \( P_1 \) and \( M_1 \), as described by the following equations:

\begin{equation}
    AR_{corr} = p_1 + p_2 \cdot \mathbb{F}
    \label{eqn:AR_corr}
\end{equation}

\begin{equation}
    \mathbb{F} = P_1^{-0.14}\sqrt{M_1}
    \label{eqn:F_old}
\end{equation}

Petersen \& Hanson \cite{Petersen_2001} suggested $p_1$ and $p_2$ to be 0.32 and 2.34, respectively, to fit their experimental data. Nativel provided two sets of values: for small diameter tubes ($p_1$ = 0.48 and $p_2$ = 0.13) and for large diameter tubes ($p_1$ = -0.53 and $p_2$ = 1.32). However, considerable scatter remained around the suggested fits. Using the same correlation, the current study adjusted $p_1$ and $p_2$ to achieve the best fit for the data. Figure \ref{fig:ISW_correlation}a shows the experimentally measured incident shock attenuation rate for the different rupture types as a function of $P_1$ and $M_1$. Equations \ref{eqn:AR_corr} and \ref{eqn:F_old} are used to fit the data, and Table \ref{tab:AR_corr} presents the values of the correlation parameters. While type-1 and type-2 ruptures show good agreement with the fit, irregular patterns seen in type-3 and type-4 ruptures exhibit significant scatter about the fit. Although the correlations capture the attenuation due to viscous effects, the contributions from the diaphragm opening process are missing. Given that earlier sections demonstrated the significant influence of the diaphragm opening process on shock tube flow, it is justified to include diaphragm opening times, shock characteristics, and opening profile parameters in predicting shock attenuation rates. Therefore, the attenuation rate is considered to be proportional to $P_1$, $M_1^{exp}$, $M_1^{idl}$, $T_{OP}$, $c_1$, $c_2$, and $c_3$. The developed correlation for the experimental data is given by the following equation:

\begin{equation}
\mathbb{F} = P_1^{-0.14} \cdot \sqrt{\frac{M_1^{exp}}{M_1^{idl}}} \cdot T_{OP}^{0.15} \cdot c_1^{-0.2} \cdot c_2^{-0.2} \cdot c_3^{0.2}
\label{eqn:F_new}
\end{equation}

It is observed that the relationship with $P_1$ and $M_1$ remains consistent with previously reported dependencies. Additionally, the experimental Mach number is normalized against the ideal Mach number, which is directly dependent on the diaphragm rupture pressure. The attenuation rate is also directly proportional to $T_{OP}$ and $c_3$. Therefore, a shorter diaphragm opening time and a rapid rupture in the initial phases of the diaphragm opening are ideal for reducing attenuation rates. It is also seen that the attenuation rate is inversely proportional to $c_1$ and $c_2$, indicating that a larger final opening aperture and a sharper slope in the diaphragm opening profile lead to lower attenuation rates. Figure \ref{fig:ISW_correlation}b shows the experimentally obtained shock attenuation rates plotted against the new correlation. Table \ref{tab:AR_corr} indicates that the $R^2$ value for the new correlation is much better than that for the previous correlation, demonstrating an improved fit and a more accurate prediction of shock attenuation rates considering the additional parameters related to the diaphragm opening process.

\begin{table}
\centering
\begin{ruledtabular}
\begin{tabular}{cccccc}
\rule{0pt}{10pt}Relation for $\mathbb{F}$ & Dependent parameters of $\mathbb{F}$ & $p_1$ & $p_2$ & $R^2$ \\ \hline
\rule{0pt}{10pt} Equation \ref{eqn:F_old} & $P_1$, $M_1^{exp}$ & 1.72 & 0.53 & 0.28 \\
\rule{0pt}{10pt} Equation \ref{eqn:F_new} & $P_1$, $M_1^{exp}$, $M_1^{idl}$, $T_{OP}$, $c_1$, $c_2$, $c_3$ & 12.44 & 0.51 & 0.63 \\    
\end{tabular}
\end{ruledtabular}
\caption{Incident shock attenuation correlation parameters for the cases investigated in this study using the equation: $AR_{corr} = p_1 + p_2 \cdot \mathbb{F}$. Here $p_1$ and $p_2$ are constants while $\mathbb{F}$ represents the expression dependent on parameters shown in the table.}
\label{tab:AR_corr}
\end{table}

\subsection{\label{sec:dpdt}Correlation for post-shock pressure rise}

Nativel et al. \cite{Nativel_2020} suggested a correlation for the post-shock pressure rise to fit their measured data of interest with independent initial parameters $P_1$, $M_1$, specific heat ratio of the driven gas ($\gamma_1$) and the internal shock tube diameter, $D$. These parameters were chosen based on their influence on the boundary-layer growth. They proposed a wide range of exponents for the parameters as their values varied based on data from different facilities. Their correlation for post-shock pressure rise, accounting for facility-dependent effects, was developed as follows,

\begin{equation}
    {\frac{dP^*}{dt}}_{corr} \propto P_1^{-0.04} M_1^{3.13} D^{-1} \gamma_1^{0.37}
    \label{eqn:nativel_dpdt}
\end{equation}

Similar to the correlation developed for incident shock wave attenuation, the equation for post-shock pressure rise is generalized here in the form of a linear correlation capturing the dependencies on $P_1$ and $M_1$. The specific heat ratio and internal diameter of the shock tube are not considered for the present fit as they remained unchanged for the experiments reported in this work. Equations \ref{eqn:DPDT_corr} and \ref{eqn:G_old} provide the generalized form based on $P_1$ and $M_1$ (mainly influenced by boundary layer effects).

\begin{equation}
    {\frac{dP^*}{dt}}_{corr} = q_1 + q_2 \cdot \mathbb{G}
    \label{eqn:DPDT_corr}
\end{equation}

\begin{equation}
    \mathbb{G} = P_1^{-0.04} M_1^{3.13}
    \label{eqn:G_old}
\end{equation}

\begin{figure*}
    \centering
    \includegraphics[width=\textwidth]{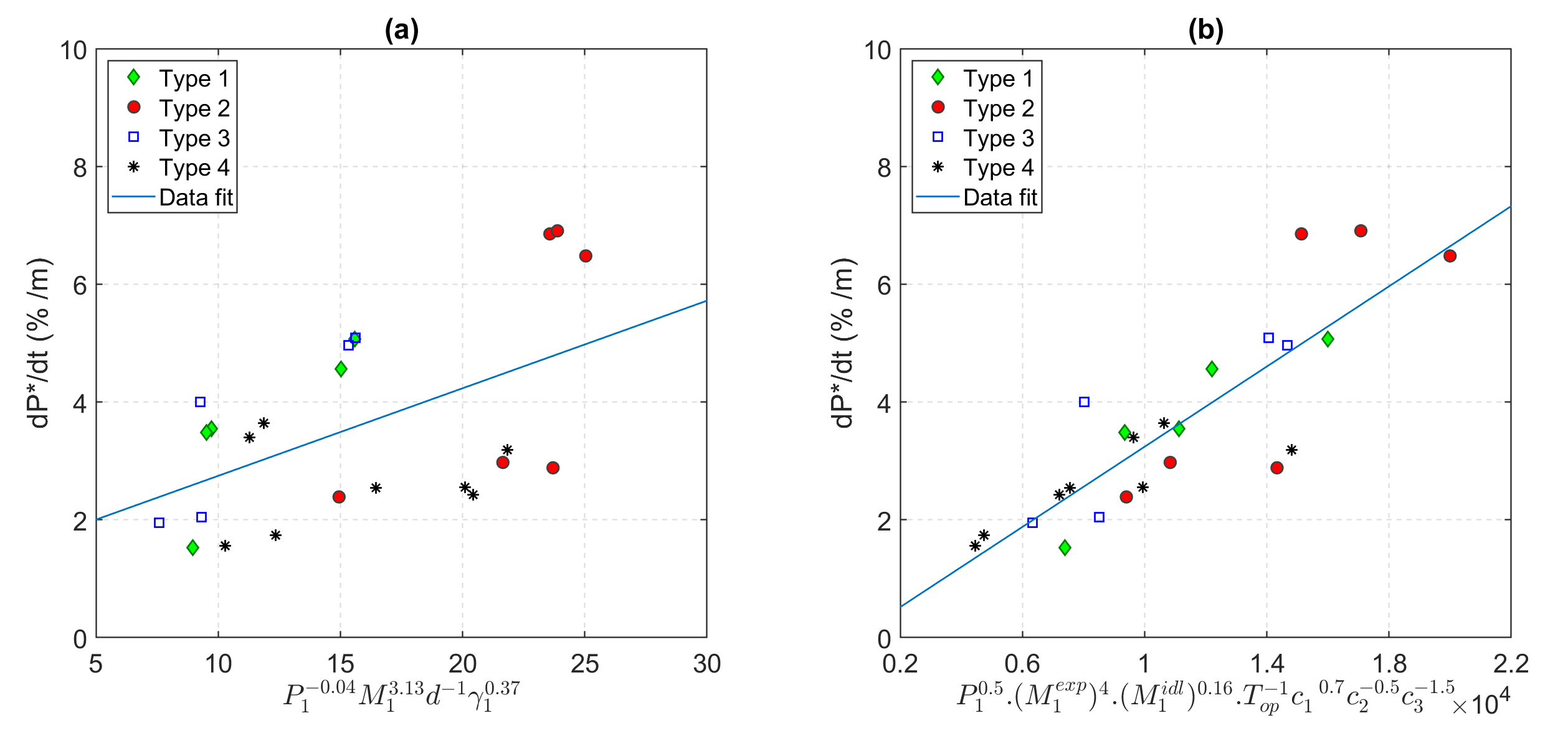}
    \caption{Comparison between existing correlation for post-shock pressure rise, d$P^*$/d$t$, and the newly developed correlation. (a) d$P^*$/d$t$ as a function of Mach number, fill pressure, shock tube diameter, and specific heat ratio\cite{Nativel_2020}. In the present experiments, the shock tube diameter and specific heat ratio is the same in all experiments. (b) d$P^*$/d$t$ as a function of shock parameters, diaphragm opening time, and opening profile parameters.}
    \label{fig:DPDT_correlation}
\end{figure*}

Using the above equations, the current study adjusted the constants $q_1$ and $q_2$ to achieve the best fit for the data. Figure \ref{fig:DPDT_correlation}a shows the experimentally measured incident shock attenuation rate for the different rupture types as a function of $P_1$ and $M_1$. Table \ref{tab:DPDT_corr} presents the values of the correlation parameters for the post-shock pressure rise. In order to include the diaphragm opening time, shock Mach number and opening profile parameters in the correlation, the post-shock pressure rise is considered to be proportional to $P_1$, $M_1^{exp}$, $M_1^{idl}$, $T_{OP}$, $c_1$, $c_2$, and $c_3$. The newly developed correlation is given by the equation:

\begin{equation}
\mathbb{G} = P_1^{0.5} \cdot (M_1^{exp})^4 \cdot (M_1^{idl})^{0.16} \cdot T_{op}^{-1} \cdot c_1^{0.7} \cdot c_2^{-0.5} \cdot c_3^{-1.5}
\label{eqn:G_new}
\end{equation}

Table \ref{tab:DPDT_corr} indicates that the $R^2$ value for the new correlation is much better than that for the previous correlation, demonstrating an improved fit and a more accurate prediction of shock attenuation rates considering the additional parameters related to the diaphragm opening process.

\begin{table}
\centering
\begin{ruledtabular}
\begin{tabular}{ccccc}
\rule{0pt}{10pt}Equation for $\mathbb{G}$ & Dependent parameters of $\mathbb{G}$ & $q_1$ & $q_2$ & $R^2$ \\ \hline
\rule{0pt}{10pt} Equation \ref{eqn:G_old} & $P_1$, $M_1^{exp}$ & 0.15 & 1.26 & 0.27 \\
\rule{0pt}{10pt} Equation \ref{eqn:G_new} & $P_1$, $M_1^{exp}$, $M_1^{idl}$, $T_{OP}$, $c_1$, $c_2$, $c_3$ & 3.4e-4 & -0.16 & 0.72 \\    
\end{tabular}
\end{ruledtabular}
\caption{Post-shock pressure rise correlation parameters for the cases investigated in this study using the equation: d$P^*$/d$t$$_{corr} = q_1 + q_2 \cdot \mathbb{G}$. Here $q_1$ and $q_2$ are constants while $\mathbb{G}$ represents the expression dependent on parameters shown in the table.}
\label{tab:DPDT_corr}
\end{table}

\section{\label{sec:concl}CONCLUSIONS}
This study extensively investigated the influence of diaphragm opening processes on shock parameters, focusing on incident shock attenuation rates and post-shock pressure rise. Various diaphragm rupture patterns were captured using high-speed imaging to classify different opening profiles and analyze their impact on shock dynamics. Four distinct types of diaphragm ruptures were identified, each demonstrating unique opening characteristics that significantly influenced the shock parameters. Type-1 ruptures exhibited a relatively symmetric and rapid opening, while type-2 ruptures were characterized by a more gradual and asymmetric opening. Type-3 ruptures showed a delayed opening with an initial pause, and type-4 ruptures displayed incomplete openings that obstructed the flow. The results show that the initial stages of diaphragm opening, including the rate and profile of opening, play crucial roles in determining the incident shock Mach number, $M_1$, and test time $\Delta t$. Type-1, type-2, and type-3 ruptures yielded \(M_1\) values closer to the ideal predictions compared to type-4 ruptures, which showed significant deviations due to incomplete diaphragm openings. The analysis also revealed that type-4 ruptures resulted in longer test times due to the delayed interaction between the reflected shock wave and the contact surface. A sigmoid function provided a generalized fit for the different rupture types. New correlations were developed to predict the incident shock attenuation rate and post-shock pressure rise, incorporating parameters such as diaphragm opening time $T_{OP}$, rupture profile constants, and normalized experimental Mach number. The new correlations demonstrated improved fits and more accurate predictions compared to previous models.

In conclusion, this study highlights the significant impact of diaphragm opening processes on shock tube flow characteristics. The findings emphasize the importance of considering the diaphragm rupture dynamics in shock tube experiments to achieve accurate predictions of shock parameters. Future work may focus on exploring different diaphragm materials and rupture mechanisms to further enhance the understanding of shock formation and attenuation in shock tubes. The analysis and prediction models will also be extended to diaphragmless shock tubes that offer several advantages over diaphragm-type shock tubes in reducing non-idealities.

\nocite{*}
\bibliography{aipsamp}

\end{document}